\documentclass[preprint,superscriptaddress,nofootinbib]{revtex4-1}
\usepackage{graphicx}
\usepackage{dcolumn}
\usepackage{bm}
\usepackage{amsmath}
\usepackage{amsfonts} 
\usepackage{latexsym}
\usepackage{bbm}
\usepackage{color}

\begin{document}

\preprint{IPMU15-0134}

\title{Electroweak interacting dark matter with a singlet scalar portal}

\author{Cheng-Wei Chiang}%
\email[e-mail: ]{chengwei@ncu.edu.tw}
\affiliation{Center for Mathematics and Theoretical Physics and Department of Physics, National Central University, Taoyuan, Taiwan 32001, R.O.C.}
\affiliation{Institute of Physics, Academia Sinica, Taipei, Taiwan 11529, R.O.C.}
\affiliation{Physics Division, National Center for Theoretical Sciences, Hsinchu, Taiwan 30013, R.O.C.}
\affiliation{Kavli IPMU (WPI), UTIAS, The University of Tokyo, Kashiwa, Chiba 277-8583, Japan}
\author{Eibun Senaha}%
\email[e-mail: ]{senaha@ncu.edu.tw}
\affiliation{Center for Mathematics and Theoretical Physics and Department of Physics, National Central University, Taoyuan, Taiwan 32001, R.O.C.}

\bigskip

\date{\today}

\begin{abstract}
We investigate an electroweak interacting dark matter (DM) model in which the DM is the neutral component of the SU$(2)_L$ triplet fermion that couples to the standard model (SM) Higgs sector via an SM singlet Higgs boson.  In this setup, the DM can have a CP-violating coupling to the singlet Higgs boson at the renormalizable level.
As long as the nonzero Higgs portal coupling (singlet-doublet Higgs boson mixing) exists,
we can probe CP violation of the DM via the electric dipole moment of the electron.
Assuming the $\mathcal{O}(1)$ CP-violating phase in magnitude, we investigate the relationship 
between the electron EDM and the singlet-like Higgs boson mass and coupling.
It is found that for moderate values of the Higgs portal couplings, current experimental EDM bound is not able to exclude the wide parameter space due to a cancellation mechanism at work.
We also study the spin-independent cross section of the DM in this model.
It is found that although a similar cancellation mechanism may diminish the leading-order correction, 
as often occurs in the ordinary Higgs portal DM scenarios, 
the residual higher-order effects leave an $\mathcal{O}(10^{-47})~{\rm cm}^2$ correction
in the cancellation region.
It is shown that our benchmark scenarios would be fully tested by combining all future experiments 
of the electron EDM, DM direct detection and Higgs physics.
\end{abstract}

\pacs{Valid PACS appear here}

\maketitle

\section{Introduction}

The existence of dark matter (DM) in the Universe is firmly established by cosmological and astronomical observations, with its relic abundance measured by the comic microwave background being~\cite{Agashe:2014kda}
\begin{align}
\Omega_{\rm CDM}h^2 = 0.1198\pm 0.0026 ~,
\end{align}
where $h$ is the reduced Hubble constant.  In spite of the undoubted existence, we still do not know where to put the DM in the particle spectrum due to the lack of solid evidence from direct searches and identification of its quantum numbers.

Although the standard model (SM) is very successful in explaining most empirical observations in particle physics, one of its shortcomings is the absence of a DM candidate.  To amend this, there have been many proposals to extend the SM with a dark sector, in which the lightest member, serving as a DM, cannot decay into SM particles due to some dark charge.

Weak-interacting massive particles (WIMP's) has attracted much attention as candidates for the DM because it is naturally accommodated in the TeV-scale physics.  For example, non-singlet DM's under the $SU(2)_L\times U(1)_Y$ emerge in supersymmetric (SUSY) models such as the minimal supersymmetric SM (MSSM) (see, {\it e.g.}, Ref.~\cite{Jungman:1995df} for a review).  On the other hand, isospin singlet DM's commonly appear in the context of the Higgs portal scenarios in which the DM's can communicate with the SM particles only via the Higgs sector~\cite{Silveira:1985rk,Burgess:2000yq,Patt:2006fw,SFDM,Baek:2011aa,Baek:2012uj,LopezHonorez:2012kv,Ghorbani:2014qpa}.  A lot of work have been done based on effective field theories or on specific renormalizable models, with both approaches complementary to each other.  The former has a strong power in probing the dark sector in a model-independent way.  However, some phenomena such as accidental cancellations due to light particles are often improperly described within this framework, and the latter is more appropriate to address such issues.

One of the unknown properties of the DM is its CP nature.  In renormalizable fermionic DM Higgs portal scenarios, it is possible for the DM to have both scalar and pseudoscalar couplings (denoted by $g^S$ and $g^P$, respectively).  Explicitly, one may have
\begin{align}
S\overline{\chi^c}\Big(g^S
	+i\gamma_5g^P\Big)\chi+{\rm h.c.} ~,\label{Schichi}
\end{align}
where $S$ an isospin singlet scalar playing the role of messenger between the dark sector and Higgs sector,
and the phase of fermionic DM field $\chi$ is already rotated so that its mass is real.  If $\chi$ is a singlet under the SM gauge symmetry, it will be hard to probe CP violation in the dark sector as its effect appears only at loop levels in the Higgs sector.  If $\chi$ participates in the electroweak interactions, on the other hand, we may detect the existence of such CP violation in electric dipole moment (EDM) experiments.

In EW-interacting DM (EWIMP) scenarios~\cite{EWIMP,MDM}, the interactions between the DM and the gauge bosons are fixed by the ordinary gauge couplings, leaving the DM mass the only unknown parameter.  However, the DM mass is also completely determined once the thermal relic scenario is assumed.  For example, the DM mass should be around 3~TeV in the Wino case~\cite{EWIMP,MDM}.  In the nonthermal relic scenario, on the other hand, 
the relic density could be explained by nonthermal production of the DM from heavier particles.  In this case, it is conceivable that the DM mass can be as light as $\mathcal{O}(100)$ GeV.

In this Letter, we consider a model in which the DM resides in an SU$(2)_L$ triplet fermion with hypercharge $Y=0$ (Wino-like DM)\footnote{Other than SUSY and inspired models, the SU$(2)_L$ triplet fermions also emerge in a specific DM model that achieves gauge coupling unification~\cite{DM_GUT}.} and the interaction given in Eq.~(\ref{Schichi}).  Here we do not confine ourself to the thermal relic scenario and, therefore, the DM mass is taken as a free parameter.  In this framework, we study the CP-violating effects coming from the dark sector on the electron EDM in connection with Higgs physics.  Throughout the analysis, the singlet scalar $S$ is assumed to be lighter than 1~TeV.  For the heavy $S$ case, the interaction between $\chi$ and Higgs doublet ($H$) would be described by the dimension-5 operator $H^\dagger H\bar{\chi^c}(g'^S+i\gamma_5 g'^P)\chi/\Lambda$ after integrating out the $S$ field.  Recent studies on the connections between CP violation and the EWIMP using the effective Lagrangian can be found in Refs.~\cite{Hisano:2014kua,Nagata:2014aoa}.

The structure of this paper is as follows.  In Section~\ref{sec:model}, we describe the DM model, with particular emphasis on the Higgs and dark sectors.  Stability and global minimum conditions for the Higgs potential are discussed.  We also provide the Higgs couplings with the SM particles and the triplet fermions.  Section~\ref{sec:pheno} discusses observables that can be used to constrain or test the model.  Numerical results of these observables are presented in Section~\ref{sec:num}.  Our findings are summarized in Section~\ref{sec:summary}.

\section{The Model}
\label{sec:model}

We consider a model in which the DM candidate arises from an SU(2)$_L$ triplet (Wino-like) fermion field $\chi$ and couples to the SM Higgs sector via an SU(2)$_L$ singlet scalar field $S$.  Both $\chi$ and $S$ are assumed to carry no hypercharge.  The relevant interactions are described by the Lagrangian
\begin{align}
\begin{split}
\mathcal{L} \supset&
\left( D_\mu H^\dagger \right)\left( D^\mu H \right) + \mu_H^2 H^\dagger H - \lambda_H |H^\dagger H|^2
+\frac{1}{2}\partial_\mu S\partial^\mu S
	+i\bar{\chi}^a\bar{\sigma}^\mu D_\mu \chi^a 
\\
&-\frac{1}{2}
\Big[
	M\chi^a\chi^a+\lambda S\chi^a\chi^a
	+\kappa\tilde{H}^\dagger\frac{\tau^a}{2}\ell_{L}\chi^a+{\rm h.c.}	
\Big]
\\
&
	-\mu_S^3S-\frac{m_S^2}{2} S^2-\frac{\mu'_S}{3}S^3-\frac{\lambda_S}{4}S^4
	-\mu_{HS}H^\dagger H S-\frac{\lambda_{HS}}{2}H^\dagger HS^2 ~,
\end{split}
\label{eq:lagrangian}
\end{align}
where $\chi^a$ denote 2-component spinors, $\tilde{H}=i\sigma^2 H^*$ and $\bar\sigma^\mu = (1,-\sigma^i)$ with $\sigma^i$ being the Pauli matrices, and the covariant derivative acting on the field $\chi^a$ is
\begin{align}
D_\mu\chi^a &= \partial_\mu\chi^a -g_2\epsilon^{abc}A_\mu^b\chi^c ~,
\end{align}
with $g_2$ being the SU(2)$_L$ gauge coupling.  We impose the $Z_2$ symmetry, $\chi\to -\chi$, so that the third term involving the lepton doublet $\ell_L$ in the square bracket of Eq.~(\ref{eq:lagrangian}) drops out, and the neutral component of $\chi$ becomes a DM candidate.  Phenomenology of DM without the singlet Higgs boson is well studied (see, for example, Refs.~\cite{EWIMP,MDM}).

We parameterize the Higgs fields as follows:
\begin{align}
H(x)=\left(
\begin{array}{c}
G^+(x) \\
\frac{1}{\sqrt{2}}\big(v+h(x)+iG^0(x)\big)
\end{array}
\right) ~,\quad 
S(x) = v_S +h_S(x) ~,
\end{align}
where $v=246$~GeV, and $G^+$ and $G^0$ are the Nambu-Goldstone bosons.  The Higgs sector of this model is the same as the real singlet-extended SM (rSM).  Here we give a quick review of rSM to make the paper self-contained.  The tadpole conditions are 
\begin{align}
\left\langle\frac{\partial V}{\partial h} \right\rangle &= v
\left[
	-\mu_H^2+\lambda_Hv^2+\mu_{HS}v_S+\frac{\lambda_{HS}}{2}v_S^2
\right] = 0 ~,
\label{treetad_h}\\
\left\langle\frac{\partial V}{\partial h_S} \right\rangle &= v_S
\left[
	\frac{\mu_S^3}{v_S}+m_S^2+\mu'_Sv_S+\lambda_Sv_S^2
	+\frac{\mu_{HS}}{2}\frac{v^2}{v_S}+\frac{\lambda_{HS}}{2}v^2
\right]=0 ~,
\label{treetad_hS}
\end{align}
where $\langle \cdots \rangle$ means that the quantity in the bracket is evaluated in the vacuum.  These two tadpole conditions can be used to solve for $\mu_H^2$ and $m_S^2$ in terms of the other parameters.
Assuming $v, v_S \neq 0$, the squared-mass matrix of the Higgs bosons in the vacuum is cast into the form
\begin{align}
\mathcal{M}_{H}^2&= 
\left(
	\begin{array}{cc}
	2\lambda_Hv^2
	& \mu_{HS}v+\lambda_{HS}vv_S \\
	\mu_{HS}v+\lambda_{HS}vv_S 
	& -\frac{\mu_S^3}{v_S}+\mu'_Sv_S+2\lambda_Sv_S^2-\frac{\mu_{HS}}{2}\frac{v^2}{v_S}
	\end{array}
\right) ~,
\label{eq:MH}
\end{align}
which can be diagonalized by an orthogonal matrix as
\begin{align}
O(\alpha)^T\mathcal{M}_H^2O(\alpha)  =
\left(
	\begin{array}{cc}
	m_{H_1}^2 & 0 \\
	0 & m_{H_2}^2
	\end{array}
\right), \quad
O(\alpha) = 
\left(
	\begin{array}{cc}
	\cos\alpha & -\sin\alpha \\
	\sin\alpha & \cos\alpha
	\end{array}
\right),
\end{align}
where $-\pi/4 \le \alpha \le \pi/4$.  Here we assume that the mass eigenvalues satisfy $m_{H_1}<m_{H_2}$, and $m_{H_1}=125$ GeV.  The scenario of no mixing between the $H$ and $S$ fields ($\alpha \to 0$) occurs in both the alignment limit $\mu_{HS} = -\lambda_{HS}v_S$ and the decoupling limit $-\frac{\mu_S^3}{v_S}+\mu'_Sv_S+2\lambda_Sv_S^2-\frac{\mu_{HS}}{2}\frac{v^2}{v_S} \gg 2\lambda_Hv^2$.

The tree-level effective potential is given by
\begin{align}
V_0(\varphi, \varphi_S) &= -\frac{\mu_H^2}{2}\varphi^2+\frac{\lambda_H}{4}\varphi^4 
	+\frac{\mu_{HS}}{2}\varphi^2\varphi_S+\frac{\lambda_{HS}}{4}\varphi^2\varphi_S^2\nonumber\\
&\quad +\mu_S^3\varphi_S+\frac{m_S^2}{2}\varphi_S^2+\frac{\mu'_S}{3}\varphi_S^3
	+\frac{\lambda_S}{4}\varphi_S^4 ~,\label{V0}
\end{align}
where $\varphi$ and $\varphi_S$ are respectively the classical background fields of $h$ and $h_S$, and $\mu_H^2$ and $m_S^2$ are given by Eqs.~(\ref{treetad_h}) and (\ref{treetad_hS}).  In order for the potential to be bounded from below, we impose the following conditions on the quartic couplings:
\begin{align}
\lambda_H>0 ~, \quad \lambda_S>0 ~,\quad -2\sqrt{\lambda_H\lambda_S}<\lambda_{HS} ~,
\end{align}
where the last condition is needed in particular when $\lambda_{HS}$ takes negative values.
Since $V_0(\varphi, \varphi_S)$ is not symmetric under the transformation $\varphi_S\to -\varphi_S$, 
it is possible for $V_0(\varphi, \varphi_S)$ 
to have another vacuum that is lower than the electroweak vacuum specified by $(v, v_S)$.
In Ref.~\cite{Baek:2012uj}, the conditions for the electroweak vacuum to be the global minimum are investigated and, as a result, it is found that
\begin{align}
\sqrt{\frac{\lambda_S}{2}}|v_S| < m_{H_2} < \sqrt{2\lambda_S}|v_S| ~,
\label{eq:global_min}
\end{align}
under the conditions $\alpha=\mu_S=0$ and $\sqrt{\lambda_{HS}}v\ll\sqrt{\lambda_S}|v_S|$.  The left inequality is derived by requiring that the electroweak vacuum has a lower energy than the symmetry vacuum, while the right inequality is obtained by demanding that the electroweak vacuum be lower than another local minimum on the $v_S$ axis.~\footnote{The existence of such a nontrivial vacuum commonly happens in the context of strong first-order electroweak phase transition, as needed for successful electroweak baryogenesis~\cite{Funakubo:2005pu,Fuyuto:2014yia}. However, the condition $\sqrt{\lambda_{HS}}v\ll\sqrt{\lambda_S}|v_S|$ usually does not hold in such cases so that the mass bound (\ref{eq:global_min}) is not valid.}
It is noted that numerically Eq.~(\ref{eq:global_min}) is still a good approximation 
even when $\alpha \sim 0.2$ [rad].  Moreover, one can turn the inequalities into
\begin{align}
\frac{m_{H_2}}{\sqrt2} < \sqrt{\lambda_S} |v_S| < \sqrt2 m_{H_2} ~.
\end{align}
Therefore, if another neutral Higgs boson is found experimentally, one can use its mass to bound $\sqrt{\lambda_S} |v_S|$ in the above-mentioned limit of the model.

Note that the constraint in Eq.~(\ref{eq:global_min}) is derived from the tree-level potential given in Eq.~(\ref{V0}).  Thus, it may change after including one-loop corrections, especially from the $\chi$-loops.  However, as long as the magnitudes of $\lambda$'s and $\alpha$ are moderate, which we assume throughout this paper, the tree-level result still remains intact.  For the explicit one-loop demonstration in the singlet fermionic DM model, see Ref.~\cite{Baek:2012uj}.

The Higgs coupling constants relevant for our analysis are
\begin{align}
\begin{split}
\mathcal{L}_{H_iVV} &= \frac{1}{v}\sum_{i=1,2}g_{H_iVV}^{}H_i(m_Z^2Z_\mu Z^\mu
+2m_W^2W_{\mu}^+ W^{-\mu}) ~, 
\\
\mathcal{L}_{H_i\bar{f}f} &= -\frac{m_f}{v}\sum_{i=1,2}g_{H_i\bar{f}f}^{}H_i\bar{f}f ~, 
\\
\mathcal{L}_{H\chi\chi} 
&=-\sum_{i=1,2}H_i\overline{\chi^+}\Big(g_{H_i\bar{\chi}\chi}^S
	+i\gamma_5g_{H_i\bar{\chi}\chi}^P\Big)\chi^+
-\frac{1}{2}\sum_{i=1,2}H_i\overline{\chi^0}\Big(g_{H_i\bar{\chi}\chi}^S
	+i\gamma_5g_{H_i\bar{\chi}\chi}^P\Big)\chi^0 ~,
\end{split}
\end{align}
where $\chi^{+(0)}$ are the 4-component Dirac (Majorana) fermions and 
\begin{align}
\begin{split}
g_{H_1VV}^{} &= g_{H_1\bar{f}f}^{}=c_\alpha,\quad g_{H_2VV}^{} = g_{H_2\bar{f}f}^{}= -s_\alpha ~, 
\\
g_{H_1\bar{\chi}\chi}^S 
	&= |\lambda|\cos\delta_\phi s_\alpha, \quad
g_{H_1\bar{\chi}\chi}^P = -|\lambda|\sin\delta_\phi s_\alpha ~,
\\
g_{H_2\bar{\chi}\chi}^S 
	&= |\lambda|\cos\delta_\phi c_\alpha, \quad
g_{H_2\bar{\chi}\chi}^P = -|\lambda|\sin\delta_\phi c_\alpha ~,
\end{split}
\label{eq:couplings}
\end{align}
with $\lambda=|\lambda|e^{i\phi_\lambda}$, $M_\chi = M+\lambda v_S = |M_\chi|e^{i\phi_{M_\chi}}$, and $\delta_\phi \equiv \phi_\lambda-\phi_{M_\chi}$ being the only physical CP-violating phase in the new sector.  Here we have also used the shorthand notations $s_\alpha = \sin\alpha$ and $c_\alpha = \cos\alpha$.  Na{\"i}vely, we expect that the phase $\delta_\phi \sim {\cal O}(1)$ and will discuss its effects in various observables.  At tree level, $\chi^\pm$ and $\chi^0$ are degenerate in mass, given by $|M_\chi|$ above.  As will be discussed in the next section, such a degeneracy is lifted by radiative corrections.  We will thus use $m_{\chi^\pm}$ and $m_{\chi^0}$ to denote the physical masses of $\chi^\pm$ and $\chi^0$, respectively.

\section{Phenomenology}
\label{sec:pheno}

Although the Higgs sector of the current DM model is virtually the same as that proposed in Refs.~\cite{SFDM,Baek:2011aa,LopezHonorez:2012kv,Baek:2012uj}, there are significant differences in certain phenomena due to the triplet fermion field $\chi^a$.  Therefore, we will focus exclusively on the observables with distinctive features in this analysis, especially those being well constrained by experiments and likely to have improvements in the near future.


In this model, the dark sector participates in electroweak interactions.  Therefore, under the assumption of a nonzero CP-violating phase, it will contribute to the EDM's of electron, neutron and atoms.  The most stringent bound of all comes from the recent experimental measurement of the thorium-monoxide EDM, which places an upper bound on the electron EDM~\cite{Baron:2013eja}:
\begin{align}
|d_e| < 8.7\times 10^{-29}~{e\;\text{cm}} \quad \text{at 90\%~C.L.} ~,
\end{align}
where $e$ denotes the electric charge of the positron.  As is well known, the two-loop Barr-Zee diagrams can have significant contributions~\cite{Barr:1990vd}.  For the electron EDM, the preponderant diagram involves the Higgs boson and photon in the loop and gives
\begin{align}
\begin{split}
\left(\frac{d_e}{e}\right)_{H\gamma}
&= \frac{\alpha_{\rm em}}{8\pi^3}\frac{m_e}{m_{\chi^\pm}v}
	\sum_ig_{H_i\bar{e}e}g_{H_i\bar{\chi}\chi}^P
	g\left(\frac{m_{\chi^\pm}^2}{m_{H_i}^2}\right)
\\
&= -\frac{\alpha_{\rm em}}{8\pi^3}\frac{m_e}{m_{\chi^\pm}v}
	|\lambda|\sin\delta_\phi s_\alpha c_\alpha
	\left[
		g\left(\frac{m_{\chi^\pm}^2}{m_{H_1}^2}\right)
		-g\left(\frac{m_{\chi^\pm}^2}{m_{H_2}^2}\right)	
	\right] ~,	
\end{split}
\end{align}
where $\alpha_{\rm em}=e^2/(4\pi)$, Eq.~(\ref{eq:couplings}) is used to obtain the second line, and
the loop function $g(\tau)$ is defined as
\begin{align}
g(\tau) &= \frac{\tau}{2}\int_0^1dx\;\frac{1}{x(1-x)-\tau}\ln\left(\frac{x(1-x)}{\tau}\right) ~.
\end{align}
In the approximation of $m_{\chi^\pm} \gg m_{H_{1,2}}$, we have
\begin{align}
\left(\frac{d_e}{e}\right)_{H\gamma}
\simeq
-\frac{\alpha_{\rm em}}{16\pi^3}\frac{m_e}{m_{\chi^\pm}v}
	|\lambda|\sin\delta_\phi s_\alpha c_\alpha
	\ln\left( \frac{m_{H_2}^2}{m_{H_1}^2} \right) ~.
\end{align}
As expected, the EDM is proportional to the sine of the CP-violating phase $\delta_\phi$.  Besides, it would be vanishing if the triplet fermion does not couple with the real scalar or in the limit of $\alpha \to 0$.  Finally, the EDM would also be suppressed if the two Higgs bosons are almost degenerate in mass, a consequence of the orthogonality of the mixing matrix $O(\alpha)$.


The spin-independent cross section of the DM with a nucleon at leading order is given by
\begin{align}
\begin{split}
\sigma_{\rm SI}(\chi^0N\to \chi^0N) 
&=
\frac{\mu_{\chi^0 N}^2m_N^2}{\pi v^2}
\left(	
	\frac{g_{H_1\bar{\chi}\chi}^S}{m_{H_1}^2}c_\alpha
	-\frac{g_{H_2\bar{\chi}\chi}^S}{m_{H_2}^2}s_\alpha
\right)^2
\left(\sum_{q=u,d,s}f_{T_q}+\frac{2}{9}f_{T_G}\right)^2 
\\
&=
\frac{\mu_{\chi^0 N}^2m_N^2}{\pi v^2}
	|\lambda|^2s_\alpha^2c_\alpha^2\cos^2\delta_\phi
\left(	
	\frac{1}{m_{H_1}^2}
	-\frac{1}{m_{H_2}^2}
\right)^2
\left(\sum_{q=u,d,s}f_{T_q}+\frac{2}{9}f_{T_G}\right)^2 ~,
\end{split}
\label{sigma_SI}
\end{align}
where $m_N$ denotes the nucleon mass, $\mu_{\chi^0N}$ is the reduced mass of the DM-nucleon system, and $f_{T_q}$ and $f_{T_G}$ are the nucleon mass fractions of quark and gluon, respectively.  In the numerical study of the DM-proton cross section, we take $f_{T_u}=0.019$, $f_{T_d}=0.027$, $f_{T_s}=0.009$, and $f_{T_G}=1-\sum_{q=u,d,s}f_{T_q}=0.945$, which are calculated in Ref.~\cite{Hisano:2012wm} based on the results of Refs.~\cite{Young:2009zb,Oksuzian:2012rzb} 
\footnote{For a recent study of $f_{T_q}$, see Refs.~\cite{Crivellin:2013ipa,Hoferichter:2015dsa}. 
We have confirmed that our numerical results of $\sigma_{\rm SI}^p$ do not change much
when using their values of $f_{T_q}$.}. 
 As mentioned above, $\sigma_{\rm SI}(\chi^0N\to \chi^0N)$ would be suppressed if $m_{H_1}\simeq m_{H_2}$, the importance of which had been emphasized in Refs.~\cite{SFDM,Baek:2011aa} (see also Refs.~\cite{Baek:2012uj,Baek:2012se}).  To have an observable cross section, we also need sufficiently large couplings between $S$ and $\chi$ and mixing between the two Higgs bosons.

In the case that the above leading-order contribution is highly suppressed, higher order effects should be taken into account. Ref.~\cite{Hisano:2011cs} has evaluated the dominant electroweak loop corrections induced by the scatterings of the EWMIP with the light quarks and gluon, assuming only one Higgs doublet of the SM.  To our knowledge, there is no such a calculation with multiple Higgs bosons, and thus more precise estimates are still unknown.  Nevertheless, as we will see in the next section, since the experimentally favored region is $\cos\alpha \gtrsim 0.95$, the singlet Higgs boson effect in our model has a suppression factor of $(1-\cos^2\alpha)\lesssim 0.1$ and is expected to be subleading.  In our numerical study, the higher-order corrections are estimated using the results of Ref.~\cite{Hisano:2011cs} as a first step toward the complete analysis.

Recently, QCD corrections up to next-to-leading order in $\alpha_s$ to $\sigma_{\rm SI}$ in the EWIMP without the singlet scalar have also been finished~\cite{Hisano:2015rsa} 
(see also Ref.~\cite{EWIMP_DD_HS}). It is found that the Wino-proton cross section $\sigma_{\rm SI}^p=2.3\times 10^{-47}~{\rm cm}^2$ for a wide mass range around 1~TeV.  It is noted, however, that if the suppression at the leading order is due to the proximity of the two Higgs mass eigenstates, the cancellation is to all orders in strong interactions.


The Higgs signal strengths are useful observables to probe the structure of the Higgs sector.  Without the dark sector, the signal strengths of $H_1$ are universally scaled by $c_\alpha^2$, provided ${\rm Br}(H_1\to H_2H_2)=0$, as assumed throughout this paper.  Once the dark sector is taken into account, however, the signal strengths are modified mainly due to the contributions of charged $\chi$ to the diphoton mode:
\begin{align}
\mu_X &\simeq \frac{\sigma(pp\to H_1\to X)}
{\sigma(pp\to H_1\to X)_{\rm SM}}
\simeq \frac{c_\alpha^4\Gamma_{\rm SM}^{\rm tot}}{\Gamma^{\rm tot}},
\quad \text{where }X = ZZ^*,~ WW^*,~ \bar{f}f ~,
\\
\Gamma^{\rm tot} 
&= c_\alpha^2\Gamma^{\rm tot}_{\rm SM}|_{\rm w/o\;\Gamma(H_1\to\gamma\gamma)}
+\Gamma(H_1\to \gamma\gamma) + \Gamma(H_1\to \chi^+\chi^-)+\Gamma(H_1\to \chi^0\chi^0) ~.
\label{eq:Gamma_toto}
\end{align}
In what follows, we assume that $\chi^\pm, \chi^0$ are sufficiently heavy so that the last two decays in Eq.~(\ref{eq:Gamma_toto}) are kinematically forbidden. 
Since the diphoton mode has a relatively small partial width, we have $\Gamma^{\rm tot}\simeq c_\alpha^2\Gamma_{\rm SM}^{\rm tot}$ and $\mu_X\simeq c_\alpha^2$ for the $ZZ^*$, $WW^*$, $\bar{f}f$ channels. 
On the other hand, the signal strength of $H_1\to \gamma\gamma$ takes the form
\begin{align}
\mu_{\gamma\gamma} 
\simeq c_\alpha^2\frac{{\rm Br}(H_1\to\gamma\gamma)}
	{{\rm Br}(H_1\to \gamma\gamma)_{\rm SM}} 
= 
\left[
	\bigg|c_\alpha+\frac{\mathcal{A}_\chi^S}{\mathcal{A}_{\rm SM}}\bigg|^2
	+\bigg|\frac{\mathcal{A}_\chi^P}{\mathcal{A}_{\rm SM}}\bigg|^2
\right]
\frac{c_\alpha^2\Gamma_{\rm SM}^{\rm tot}}{\Gamma^{\rm tot}} ~,
\label{mugamgam}
\end{align}
where $\mathcal{A}_{\rm SM}=-6.49$~\cite{McKeen:2012av}, $\Gamma_{\rm SM}^{\rm tot}\simeq4.1$ MeV~\cite{Heinemeyer:2013tqa}, and 
\begin{align}
\mathcal{A}_\chi^S 
&= \frac{vg_{H_1\bar{\chi}\chi}^S}{m_{\chi^\pm}}2\tau_\chi
\big\{1+(1-\tau_\chi)f(\tau_\chi)\big\} ~, \quad
\mathcal{A}_\chi^P
= \frac{vg_{H_1\bar{\chi}\chi}^P}{m_{\chi^\pm}}2\tau_\chi f(\tau_\chi) ~,
\end{align}
with $\tau_\chi = 4m_{\chi^\pm}^2/m_{H_1}^2$ and the loop function $f(\tau_\chi)$ defined 
in Ref.~\cite{Gunion:1989we}.
In the limits of small $\alpha$ and large $m_{\chi^\pm}$, $\mu_{\gamma\gamma}$ reduces to
\begin{align}
\mu_{\gamma\gamma}\simeq 
c_\alpha^2\left[1+\frac{8v}{3m_{\chi^\pm}\mathcal{A}_{\rm SM}}
|\lambda|\cos\delta_\phi t_\alpha\right] ~,
\label{mugamgam_app}
\end{align}
where terms of higher order in $t_\alpha$ and $v/m_{\chi^\pm}$ have been neglected.
Therefore, $\mu_{\gamma\gamma} ~(\sim c_\alpha^2)$ would be suppressed in this limit.
However, it should be stressed that the reduction factor differs from both the Higgs portal DM models, such as those in Refs.~\cite{Baek:2011aa,Baek:2012uj,Baek:2012se}, and the Wino DM case in the EWIMP scenarios~\cite{EWIMP,MDM}.
Since the CP-violating part does not interfere with the SM contribution, as seen in Eq.~(\ref{mugamgam}), 
its effect is higher order in powers of $t_\alpha$ and $v/m_{\chi^\pm}$.

\section{Numerical analysis}
\label{sec:num}

Before presenting numerical results, we first summarize some current experimental constraints.
The current LHC data constrain the Higgs boson couplings of $H_1$ as~\cite{Aad:2014eha,Aad:2015gba,CMS:utj,Khachatryan:2014ira}
\begin{align}
\kappa_V &= 1.09\pm0.07~(\text{ATLAS}),\quad 
\kappa_V = 1.01^{+0.07}_{-0.07}~(\text{CMS}), \\
\kappa_F &= 1.11\pm0.16~(\text{ATLAS}),\quad 
\kappa_F = 0.89^{+0.14}_{-0.13}~(\text{CMS}), \\
\mu_{\gamma\gamma} &= 1.17\pm 0.27~(\text{ATLAS}),\quad 
\mu_{\gamma\gamma} = 1.14^{+0.26}_{-0.23}~(\text{CMS}).
\end{align}
In this model, $\kappa_V=g_{H_1VV}=c_\alpha$, $\kappa_F=g_{H_1\bar{f}f}=c_\alpha$.  Furthermore, direct searches of the second Higgs boson have been conducted using the diboson decay modes, and $m_{H_2}$ is bounded as a function of $\sin^2\alpha$ with $\mathcal{B}_{\rm new}=0.0$, $0.2$ and $0.5$, where $\mathcal{B}_{\rm new}$ denotes the contribution to the Higgs boson width from non-SM decays~\cite{Khachatryan:2015cwa}.  In the current analysis, we take $\mathcal{B}_{\rm new}=0.0$ in order to impose a most conservative constraint on $m_{H_2}$ and $\alpha$.

Direct searches of DM through spin-independent interactions have been carried out in many experiments, restraining possible DM-nucleon scattering cross section $\sigma_{\rm SI}$ over a wide range of mass, from a few GeV to TeV.  Currently, the strongest bound on $\sigma_{\rm SI}$ comes from the LUX experiment~\cite{Akerib:2013tjd}.
For instance, $\sigma_{\rm SI}\lesssim 4.7~(33)\times 10^{-45}~{\rm cm}^2$ for $m_{\chi^0}=400~(2900)$ GeV.

With a mild dependence on $M_\chi$, the mass difference $\Delta M \sim {\cal O}(100~\text{MeV})$~\cite{Ibe:2012sx,Yamada:2009ve}.  As a consequence, $\chi^\pm$ have a relatively long lifetime of $\mathcal{O}(0.1)$ ns, with the dominant decay mode of $\chi^\pm\to \pi^\pm\chi^0$.  We can probe such a meta-stable particle at colliders by identifying the disappearance of a charged track.  The ATLAS Collaboration has put a constraint on such a long-lived charged particle.  With the LHC Run-1 data, the lower bound of $m_{\chi^0}$ is found to be~\cite{Aad:2013yna}
\begin{align}
m_{\chi^0} > 270~\text{GeV\quad (95\% CL).}
\end{align}

The constraints coming from the cosmic rays are also important.  Ref.~\cite{Bhattacherjee:2014dya} analyzed the observations of gamma-rays from classical dwarf spheroidal galaxies, and found that 
\begin{align}
320~\text{GeV}\lesssim m_{\chi^0} \lesssim 2250~\text{GeV} ~,\quad
2430~\text{GeV}\lesssim m_{\chi^0} ~,
\label{eq:ranges}
\end{align}
and $m_{\chi^0}\lesssim 2900~\text{GeV}$ from the DM relic abundance constraint.

In the following analysis, we also regard the case of $m_{\chi^0}=2900$ GeV 
as the thermal relic scenario inferred by the Wino DM case.
This holds as long as the coupling between $\chi$ and $S$ is smaller than the gauge couplings.
If this is not the case, the DM mass might be changed due to the additional Sommerfeld effect induced
by $S$.
Although it is interesting to investigate such a case, the detailed analysis leaves the main scope 
of this Letter.
Throughout our analysis, we take $|\lambda|=0.1$ and focus on $\sin\delta_\phi\ge1/\sqrt{2}$,
which yields $g_{H_2\bar{\chi}\chi}^S\lesssim0.07$.

\begin{figure}[t]
\center
\includegraphics[width=0.45\linewidth]{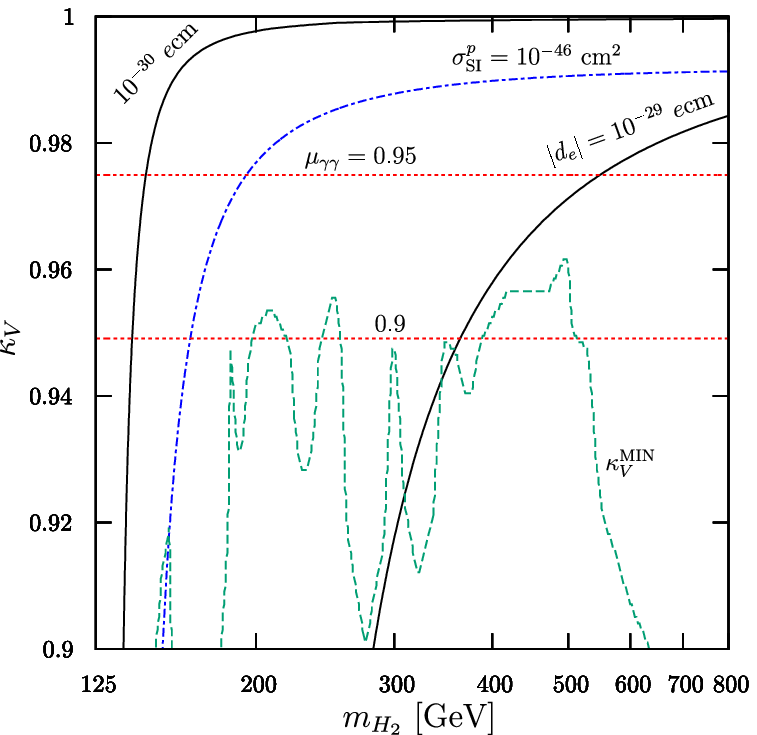}
\hspace{5mm}
\includegraphics[width=0.45\linewidth]{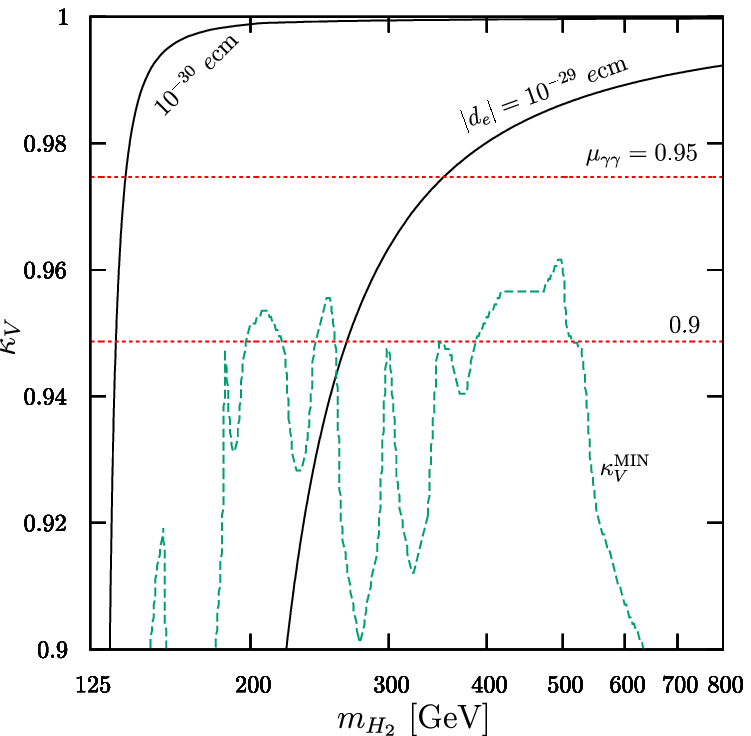}
\caption{Constraints and predictions of observables on the $m_{H_2}$-$\kappa_V$ plane.  The green dashed curve is for the lower bound on $\kappa_V$; the red dotted lines for $\mu_{\gamma\gamma}$; the black solid curves for $|d_e|$; and the blue dot-dashed curve for $\sigma_{\rm SI}^p$. We take $m_{\chi^0}=2900$~GeV, $|\lambda| = 0.1$, and $\delta_\phi=45^\circ$ (left) or $\delta_\phi=90^\circ$ (right).  In the right plot, we have $\sigma_{\rm SI}^p\simeq 1.5\times 10^{-47}~{\rm cm}^2$ in the entire region, which is the loop corrections.}
\label{fig:mH2_kv_M2900}
\end{figure}

We first present the results for $m_{\chi^0}=2900$~GeV.  In Fig.~\ref{fig:mH2_kv_M2900}, $\mu_{\gamma\gamma}$, $|d_e|$ and $\sigma_{\rm SI}^p$ are shown in the $(m_{H_2}, \kappa_V)$ plane, taking $\delta_\phi=45^\circ$ (left) and $90^\circ$ (right).  The green dashed curve gives the lower bound on $\kappa_V$ obtained by CMS with $\mathcal{B}_{\rm new} = 0$~\cite{Khachatryan:2015cwa}.  When one takes a finite value for $\mathcal{B}_{\rm new}$, the curve will shift downwards.
The red dotted lines represent $\mu_{\gamma\gamma}=0.95$ (top) and 0.9 (bottom).  Since the effects of $\chi^\pm$ are substantially decoupled, the deviation of $\mu_{\gamma\gamma}$ is virtually due to $c_\alpha^2~(=\kappa_V^2)$, as seen in Eq.~(\ref{mugamgam_app}).
The contours of electron EDM are displayed by the black solid lines: $|d_e|=10^{-29}~e\;\text{cm}$ and $10^{-30}~e~\text{cm}$ from bottom to top.  The current bound is outside the region.  As discussed above, the cancellation between $H_1$ and $H_2$ corrections gets more prominent as $m_{H_2}$ approaches 125 GeV.  Therefore, the maximal CP violation case is still allowed even if the electron EDM is improved to $10^{-30}~e~\text{cm}$.
We emphasize that this possibility cannot be encoded in the effective field theory approach
as mentioned in Introduction.

For the $\delta_\phi=45^\circ$ case, the contour of the DM direct detection cross section $\sigma_{\rm SI}^p=10^{-46}~{\rm cm}^2$ is also shown by the blue dot-dashed curve.  
Similar to the election EDM, the cancellation mechanism is at work when $m_{H_1}\simeq m_{H_2}$.
Note that even if the leading contribution in $\sigma_{\rm SI}^p$ vanishes, 
the NLO contribution ($\sigma_{\rm SI}^p\simeq 1.5\times 10^{-47}~{\rm cm}^2$) still remains, 
which yields the minimum value in the region we are considering here.
For the $\delta_\phi=90^\circ$ case, on the other hand, there is no leading-order correction 
since $\sigma_{\rm SI}^p\propto \cos^2\delta_\phi$, as shown in Eq.~(\ref{sigma_SI}).
In this case, we have $\sigma_{\rm SI}^p\simeq 1.5\times 10^{-47}~{\rm cm}^2$ in the entire region.

\begin{figure}[t]
\center
\includegraphics[width=0.45\linewidth]{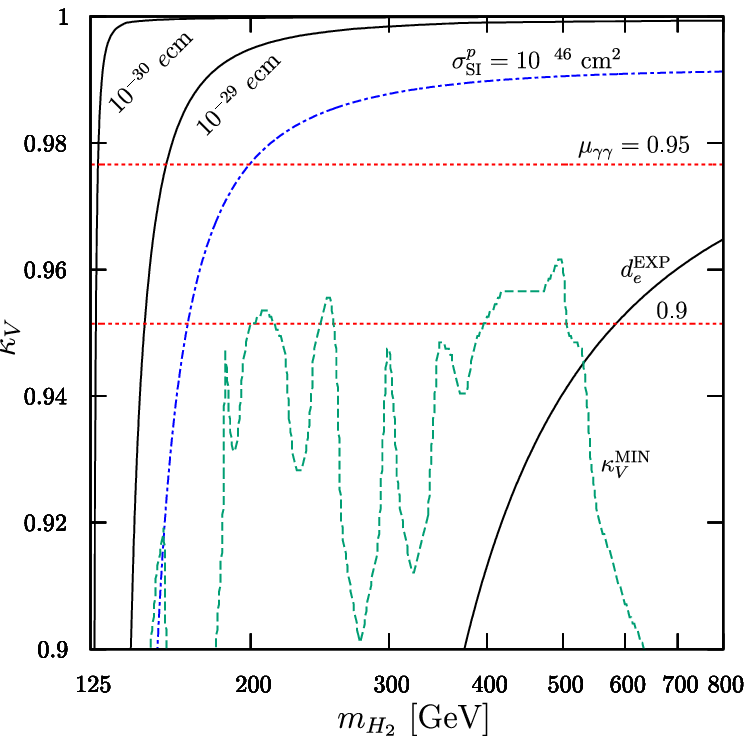}
\hspace{5mm}
\includegraphics[width=0.45\linewidth]{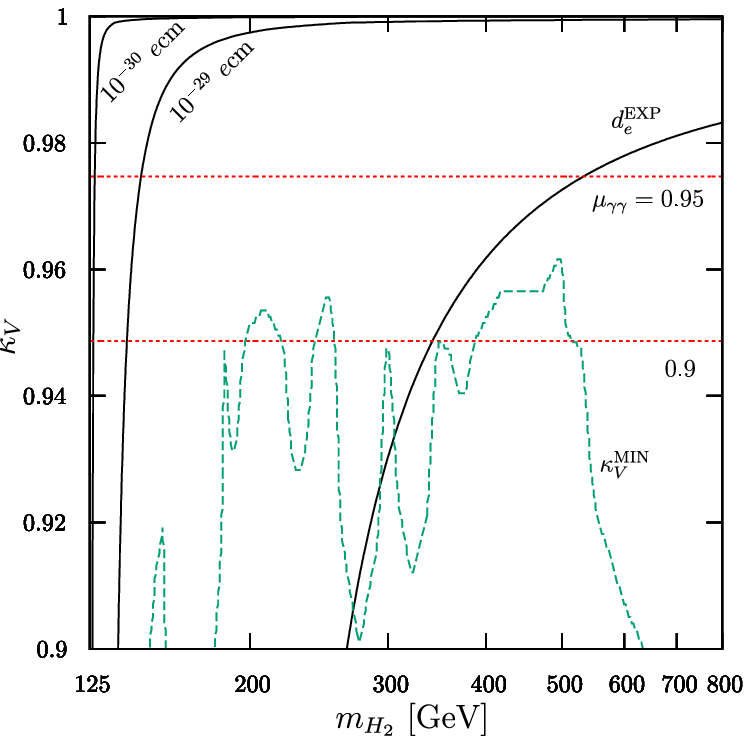}
\caption{Same as Fig.~\ref{fig:mH2_kv_M2900}, but with $m_{\chi^0}=400$~GeV.}
\label{fig:mH2_kv_M400}
\end{figure}

The results for $m_{\chi^0}=400$~GeV are given in Fig.~\ref{fig:mH2_kv_M400}, with no change in the green dashed curve and red dotted lines from Fig.~\ref{fig:mH2_kv_M2900}.  The change in the blue dot-dashed curve is tiny because the dependence of $m_{\chi^0}$ enters via the reduced mass $\mu_{\chi^0N}$.  However, the current experimental bound, indicated by the curve labeled $d_e^{\text{EXP}}$, has ruled out the parameter space below it at $90\%$ CL.
Nevertheless, we point out that a substantially large region is still viable owing to the cancellation mechanism.

\begin{figure}[t]
\center
\includegraphics[width=0.45\linewidth]{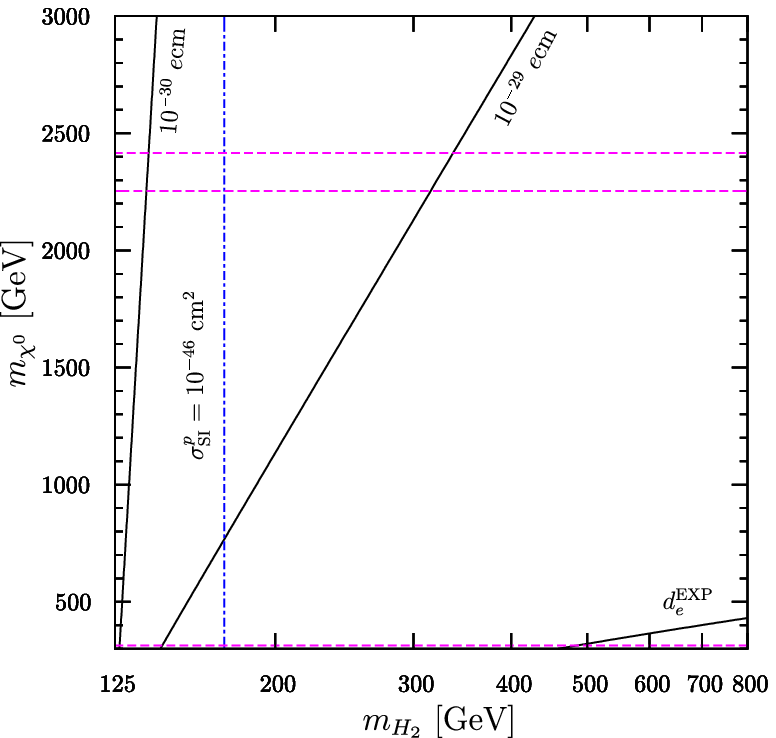}
\hspace{5mm}
\includegraphics[width=0.45\linewidth]{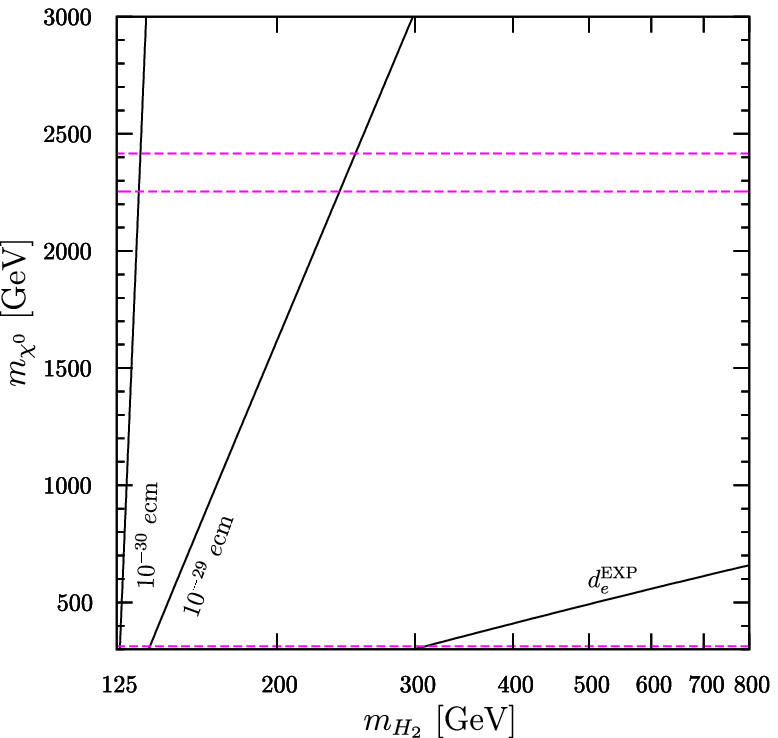}
\caption{Contours of electron EDM (solid black lines), spin-independent DM-nucleon scattering cross section (blue dot-dashed line), and allowed DM mass ranges in the plane of $m_{H_2}$ and $m_{\chi^0}$, taking $|\lambda| = 0.1$, $\cos\alpha = 0.96$, and $\delta_\phi=45^\circ$ (left) or $\delta_\phi=90^\circ$ (right).}
\label{fig:mH2_Mchi}
\end{figure}

So far, we have focused on $m_{\chi^0}=2900$~GeV and 400~GeV as two benchmark values.
Let us now consider the other cases of $m_{\chi^0}$.
In Fig.~\ref{fig:mH2_Mchi}, we show the contours of $|d_e|$ (solid black lines), $\sigma_{\text{SI}}^p$ (blue dot-dashed line), and allowed DM mass ranges (magenta dashed lines) given in Eq.~(\ref{eq:ranges}) on the $m_{H_2}$-$m_{\chi^0}$ plane, again taking $\delta_\phi=45^\circ$ (left) and $90^\circ$ (right).
From these plots, one can find that the electron EDM has the specific dependences of $m_{H_2}$
and $m_{\chi^0}$, and the patterns of which are the unique characterization of this DM model.
We here note that if the experimental bound is improved to $|d_e|=10^{-30}~e~\text{cm}$,
the only possible value of $m_{H_2}$ is around 125 GeV, and its sensitivity to $m_{\chi^0}$ is lost.

For the DM direct detection, the contour of $\sigma_{\rm SI}^p=10^{-46}~{\rm cm}^2$
is given in the case of $\delta_\phi=45^\circ$.
As seen, $\sigma_{\rm SI}^p$ is not sensitive to the DM mass
since the leading contribution in $\sigma_{\rm SI}^p$ is mostly controlled 
by $\alpha$, $g_{H_1\bar{\chi}\chi}^S$, $g_{H_2\bar{\chi}\chi}^S$ and $m_{H_2}$,
and the $m_{\chi^0}$ dependence enters only via $\mu_{\chi^0N}$, as mentioned above.

Before closing this section, a few remarks about future prospects are in order.
The improvements in the bounds on $\kappa_V$, $\mu_{\gamma\gamma}$ and $m_{H_2}$ 
are in progress at LHC Run-2, and will continue in future collider experiments,
such as the high-luminosity LHC~\cite{HL-LHC}, International Linear Collider~\cite{ILC}
and TLEP~\cite{Gomez-Ceballos:2013zzn}.
For instance, the sensitivity of $\kappa_V$ is expected to be improved up to $\mathcal{O}(0.1)$\%
at the latter two lepton colliders.

The projected sensitivity of the electron EDM in future experiments 
is around $10^{-30}~e~\text{cm}$~\cite{future_eEDM}.
In addition to this, the EDMs of nucleons and atoms may also be important 
(for a recent review, see, {\it e.g.}, Ref.~\cite{Engel:2013lsa}).

Several DM direct detection experiments are also planned.
The XENON1T experiment~\cite{Aprile:2012zx} has a better sensitivity than the current LUX bounds
by more than an order of magnitude, {\it i.e.},  
$\sigma_{\rm SI}^p=(1.2 - 50)\times10^{-47}~{\rm cm}^2$ for the DM mass in the range of $100-3000$~GeV,
which may be further improved to 
$\sigma_{\rm SI}^p=(1.8-48)\times10^{-48}~{\rm cm}^2$ by the LZ experiment~\cite{Cushman:2013zza}. 

In summary, the entire region for our benchmark points will be fully testable in these future experiments.

\section{Conclusions}
\label{sec:summary}

We have studied the phenomenology in the electroweak-interacting fermionic dark matter (DM) 
with a singlet scalar portal model.
The DM is the neutral component of the SU$(2)_L$ triplet fermion (Wino-like DM)
that has both scalar and pseudoscalar
couplings to the standard model (SM) singlet Higgs boson.  Therefore, CP symmetry can be violated at the renormalizable level.
As long as the singlet-doublet Higgs bosons mixing is nonzero,
such a CP violating effect is manifest in the visible sector.
We have investigated the relationship between the electron EDM and the singlet-like Higgs boson
mass and coupling, with and without the thermal relic scenario:
$m_{\chi^0}=$ 2900 GeV and 400 GeV, respectively.
It is found that an $\mathcal{O}(1)$ CP-violating phase is still possible on account of the cancellation 
between the two Higgs boson contributions when the two masses are close to each other.

We have also considered the direct detection bounds on the DM and found that 
if $m_{H_2}\gtrsim150$ GeV and $\kappa_V\lesssim 0.99$, the spin-independent DM-nucleon cross section $\sigma_{\rm SI}^p\simeq \mathcal{O}(10^{-46})~{\rm cm}^2$.
Therefore, the upcoming XENON1T experiment can readily probe such a region.
For $m_{H_1}\simeq m_{H_2}$, on the other hand, the cancellation mechanism is effective so that the leading-order contribution vanishes, as observed in the ordinary Higgs portal DM scenarios.
Nevertheless, since the DM participates in electroweak interactions in our model, the residual higher-order corrections still remain and amount to $\sigma_{\rm SI}^p\simeq 1.5\times 10^{-47}~{\rm cm}^2$.
The current analysis have shown that our benchmark scenario will be entirely tested by the future experiments of the electron EDM, DM direct detection and Higgs physics.

Finally, we summarize by pointing out distinctive features of our model in comparison with two existing ones.
In the model studied in Ref.~\cite{Hisano:2014kua}, the Wino DM couples with the SM Higgs boson via the dimension-5 operator $H^\dagger H\bar{\chi^c}(g^S+i\gamma_5 g^P)\chi/\Lambda$, with $\Lambda$ being a heavy mass scale.  Within the effective field theory framework, the regime with accidental cancellation, as explicitly shown in this Letter, is not properly treated.  Therefore, the two models will have different signals in CP violation associated with Higgs physics.
In the SU$(2)_L$ singlet fermionic DM model~\cite{LopezHonorez:2012kv}, the Higgs signal strengths are almost the same as those in our model.  Even though its DM sector can also accommodate CP violation, the manifestation is so dim that the electron EDM is far below the detectable level, a clear difference from our model.

\appendix

\begin{acknowledgments}
The authors would like to thank Junji Hisano, Shigeki Matsumoto, Natsumi Nagata and Sming Tsai for useful conversations.  This work was supported in part by the Ministry of Science and Technology of Taiwan under Grant Nos. MOST-100-2628-M-008-003-MY4, 104-2628-M-008-004-MY4, 104-2811-M-008-011, and 104-2811-M-008-056, and in part by the World Premier International Research Center Initiative (WPI), MEXT, Japan.
\end{acknowledgments}



\begin{thebibliography}{99}
\bibitem{Agashe:2014kda}
  K.~A.~Olive {\it et al.}  [Particle Data Group Collaboration],
  Chin.\ Phys.\ C {\bf 38} (2014) 090001.

\bibitem{Jungman:1995df} 
  G.~Jungman, M.~Kamionkowski and K.~Griest,
  Phys.\ Rept.\  {\bf 267}, 195 (1996).

\bibitem{Silveira:1985rk} 
  V.~Silveira and A.~Zee,
  Phys.\ Lett.\ B {\bf 161}, 136 (1985).

\bibitem{Burgess:2000yq}
  C.~P.~Burgess, M.~Pospelov and T.~ter Veldhuis,
  Nucl.\ Phys.\ B {\bf 619} (2001) 709.

\bibitem{Patt:2006fw} 
  B.~Patt and F.~Wilczek,
  hep-ph/0605188.

\bibitem{SFDM}  
  Y.~G.~Kim, K.~Y.~Lee and S.~Shin,
  JHEP {\bf 0805} (2008) 100.
See also, 
  Y.~G.~Kim and S.~Shin,
  JHEP {\bf 0905} (2009) 036.
  
\bibitem{Baek:2011aa}
  S.~Baek, P.~Ko and W.~I.~Park,
  JHEP {\bf 1202} (2012) 047.

\bibitem{Baek:2012uj}
  S.~Baek, P.~Ko, W.~I.~Park and E.~Senaha,
  JHEP {\bf 1211} (2012) 116.

\bibitem{LopezHonorez:2012kv}
  L.~Lopez-Honorez, T.~Schwetz and J.~Zupan,
  Phys.\ Lett.\ B {\bf 716} (2012) 179.

\bibitem{Ghorbani:2014qpa}
  K.~Ghorbani,
  JCAP {\bf 1501} (2015) 015.

\bibitem{EWIMP}
  J.~Hisano, S.~Matsumoto and M.~M.~Nojiri,
  Phys.\ Rev.\ D {\bf 67} (2003) 075014;~
%
  Phys.\ Rev.\ Lett.\  {\bf 92} (2004) 031303;~
%
  J.~Hisano, S.~Matsumoto, M.~M.~Nojiri and O.~Saito,
  Phys.\ Rev.\ D {\bf 71} (2005) 015007;~
%
  Phys.\ Rev.\ D {\bf 71} (2005) 063528;~
%
  J.~Hisano, S.~Matsumoto, O.~Saito and M.~Senami,
  Phys.\ Rev.\ D {\bf 73} (2006) 055004;~
%
  J.~Hisano, S.~Matsumoto, M.~Nagai, O.~Saito and M.~Senami,
  Phys.\ Lett.\ B {\bf 646} (2007) 34.

\bibitem{MDM}
  M.~Cirelli, N.~Fornengo and A.~Strumia,
  Nucl.\ Phys.\ B {\bf 753} (2006) 178;~
  M.~Cirelli, A.~Strumia and M.~Tamburini,
  Nucl.\ Phys.\ B {\bf 787} (2007) 152.

\bibitem{DM_GUT}
  K.~Kainulainen, K.~Tuominen and J.~Virkajarvi,
  Phys.\ Rev.\ D {\bf 82} (2010) 043511;~
  K.~Kainulainen, K.~Tuominen and J.~Virkajarvi,
  JCAP {\bf 1310} (2013) 036;~
  K.~Kainulainen, K.~Tuominen and J.~Virkajarvi,
  JCAP {\bf 1507} (2015) 07,  034.

\bibitem{Hisano:2014kua}
  J.~Hisano, D.~Kobayashi, N.~Mori and E.~Senaha,
  Phys.\ Lett.\ B {\bf 742} (2015) 80.

\bibitem{Nagata:2014aoa}
  N.~Nagata and S.~Shirai,
  Phys.\ Rev.\ D {\bf 91} (2015) 5,  055035.

\bibitem{Funakubo:2005pu}
  K.~Funakubo, S.~Tao and F.~Toyoda,
  Prog.\ Theor.\ Phys.\  {\bf 114} (2005) 369.

\bibitem{Fuyuto:2014yia}
  K.~Fuyuto and E.~Senaha,
  Phys.\ Rev.\ D {\bf 90} (2014) 1,  015015.

\bibitem{Baron:2013eja}
  J.~Baron {\it et al.} [ACME Collaboration],
  Science {\bf 343} (2014) 269.

\bibitem{Barr:1990vd}
  S.~M.~Barr and A.~Zee,
  Phys.\ Rev.\ Lett.\  {\bf 65} (1990) 21
   [Phys.\ Rev.\ Lett.\  {\bf 65} (1990) 2920].

\bibitem{Hisano:2012wm}
  J.~Hisano, K.~Ishiwata and N.~Nagata,
  Phys.\ Rev.\ D {\bf 87} (2013) 035020.

\bibitem{Young:2009zb}
  R.~D.~Young and A.~W.~Thomas,
  Phys.\ Rev.\ D {\bf 81} (2010) 014503.
  
\bibitem{Oksuzian:2012rzb}
  H.~Ohki {\it et al.} [JLQCD Collaboration],
  Phys.\ Rev.\ D {\bf 87} (2013) 034509.
  
\bibitem{Crivellin:2013ipa}
  A.~Crivellin, M.~Hoferichter and M.~Procura,
  Phys.\ Rev.\ D {\bf 89} (2014) 054021.

\bibitem{Hoferichter:2015dsa}
  M.~Hoferichter, J.~Ruiz de Elvira, B.~Kubis and U.~G.~Mei$\ss$ner,
  arXiv:1506.04142 [hep-ph].
  
\bibitem{Baek:2012se}
  S.~Baek, P.~Ko, W.~I.~Park and E.~Senaha,
  JHEP {\bf 1305} (2013) 036.

\bibitem{Hisano:2011cs}
  J.~Hisano, K.~Ishiwata, N.~Nagata and T.~Takesako,
  JHEP {\bf 1107} (2011) 005.

\bibitem{Hisano:2015rsa}
  J.~Hisano, K.~Ishiwata and N.~Nagata,
  JHEP {\bf 1506} (2015) 097.
  
\bibitem{EWIMP_DD_HS}
  R.~J.~Hill and M.~P.~Solon,
  Phys.\ Rev.\ Lett.\  {\bf 112} (2014) 211602;~
%
  R.~J.~Hill and M.~P.~Solon,
  Phys.\ Rev.\ D {\bf 91} (2015) 043504;~
  R.~J.~Hill and M.~P.~Solon,
  Phys.\ Rev.\ D {\bf 91} (2015) 043505.
  
\bibitem{McKeen:2012av} 
  D.~McKeen, M.~Pospelov and A.~Ritz,
  Phys.\ Rev.\ D {\bf 86}, 113004 (2012).

\bibitem{Heinemeyer:2013tqa} 
  S.~Heinemeyer {\it et al.}  [LHC Higgs Cross Section Working Group Collaboration],
  arXiv:1307.1347 [hep-ph].

\bibitem{Gunion:1989we}
  J.~F.~Gunion, H.~E.~Haber, G.~L.~Kane and S.~Dawson,
  Front.\ Phys.\  {\bf 80} (2000) 1.

\bibitem{Aad:2014eha}
  G.~Aad {\it et al.} [ATLAS Collaboration],
  Phys.\ Rev.\ D {\bf 90} (2014) 11,  112015.

\bibitem{Aad:2015gba}
  G.~Aad {\it et al.} [ATLAS Collaboration],
  arXiv:1507.04548 [hep-ex].

\bibitem{CMS:utj}
  [CMS Collaboration],
  CMS-PAS-HIG-13-004.

\bibitem{Khachatryan:2014ira}
  V.~Khachatryan {\it et al.} [CMS Collaboration],
  Eur.\ Phys.\ J.\ C {\bf 74} (2014) 10,  3076.
  
\bibitem{Khachatryan:2015cwa}
  V.~Khachatryan {\it et al.} [CMS Collaboration],
  arXiv:1504.00936 [hep-ex].
  
\bibitem{Akerib:2013tjd}
  D.~S.~Akerib {\it et al.} [LUX Collaboration],
  Phys.\ Rev.\ Lett.\  {\bf 112} (2014) 091303.
    
\bibitem{Ibe:2012sx}
  M.~Ibe, S.~Matsumoto and R.~Sato,
  Phys.\ Lett.\ B {\bf 721} (2013) 252.
 
\bibitem{Yamada:2009ve}
  Y.~Yamada,
  Phys.\ Lett.\ B {\bf 682} (2010) 435.
     
\bibitem{Aad:2013yna}
  G.~Aad {\it et al.} [ATLAS Collaboration],
  Phys.\ Rev.\ D {\bf 88} (2013) 11,  112006.

\bibitem{Bhattacherjee:2014dya}
  B.~Bhattacherjee, M.~Ibe, K.~Ichikawa, S.~Matsumoto and K.~Nishiyama,
  JHEP {\bf 1407} (2014) 080.

\bibitem{HL-LHC}
  [ATLAS Collaboration],
  arXiv:1307.7292 [hep-ex];~
%
  [CMS Collaboration],
  arXiv:1307.7135.

\bibitem{ILC}
  H.~Baer, T.~Barklow, K.~Fujii, Y.~Gao, A.~Hoang, S.~Kanemura, J.~List and H.~E.~Logan {\it et al.},
  arXiv:1306.6352 [hep-ph];~
%
  T.~Barklow, J.~Brau, K.~Fujii, J.~Gao, J.~List, N.~Walker and K.~Yokoya,
  arXiv:1506.07830 [hep-ex].

\bibitem{Gomez-Ceballos:2013zzn}
  M.~Bicer {\it et al.} [TLEP Design Study Working Group Collaboration],
  JHEP {\bf 1401} (2014) 164.
 
\bibitem{future_eEDM}
  Y.~Sakemi, K.~Harada, T.~Hayamizu, M.~Itoh, H.~Kawamura, S.~Liu, H.~S.~Nataraj and A.~Oikawa {\it et al.},
  J.\ Phys.\ Conf.\ Ser.\  {\bf 302} (2011) 012051;~
%
  D.~M.~Kara, I.~J.~Smallman, J.~J.~Hudson, B.~E.~Sauer, M.~R.~Tarbutt and E.~A.~Hinds,
  New J.\ Phys.\  {\bf 14} (2012) 103051;~
%
  D.~Kawall,
  J.\ Phys.\ Conf.\ Ser.\  {\bf 295} (2011) 012031.

\bibitem{Engel:2013lsa}
  J.~Engel, M.~J.~Ramsey-Musolf and U.~van Kolck,
  Prog.\ Part.\ Nucl.\ Phys.\  {\bf 71} (2013) 21.
  
\bibitem{Aprile:2012zx}
  E.~Aprile [XENON1T Collaboration],
  Springer Proc.\ Phys.\  {\bf 148} (2013) 93.

\bibitem{Cushman:2013zza}
  P.~Cushman, C.~Galbiati, D.~N.~McKinsey, H.~Robertson, T.~M.~P.~Tait, D.~Bauer, A.~Borgland and B.~Cabrera {\it et al.},
  arXiv:1310.8327 [hep-ex].

\end{thebibliography}
\end{document}